\documentclass[conference]{IEEEtran}
\IEEEoverridecommandlockouts
\usepackage{cite}
\usepackage{enumitem} 
\usepackage{amsmath,amssymb,amsfonts}
\usepackage{algorithmic}
\usepackage{graphicx}
\usepackage{hyperref}
\usepackage{textcomp}
\usepackage{xcolor}
\def\BibTeX{{\rm B\kern-.05em{\sc i\kern-.025em b}\kern-.08em
    T\kern-.1667em\lower.7ex\hbox{E}\kern-.125emX}}

\usepackage{tikz}
\usepackage{amsmath}

\usepackage{booktabs}
\usepackage{tabularx}
\newtheorem{definition}{Definition}

\usepackage{multirow}

\usepackage{etoolbox}
\patchcmd{\itemize}
  {\def\makelabel}
  {\setlength{\leftmargin}{1.5em}\def\makelabel}
  {}{}

\patchcmd{\enumerate}
  {\def\makelabel}
  {\setlength{\leftmargin}{1.5em}\def\makelabel}
  {}{}
    
\begin{document}

\title{In-Memory Indexing and Querying of Provenance in Data Preparation Pipelines
\thanks{This work received support from the National Research Agency under the France 2030
program, with reference to ANR-22-PESN-0007.}
}

\author{
\IEEEauthorblockN{Khalid Belhajjame}
\IEEEauthorblockA{
\textit{LAMSADE, Univ. Paris-Dauphine, PSL} \\
Paris, France \\
khalid.belhajjame@lamsade.dauphine.fr}
\and
\IEEEauthorblockN{Haroun Mezrioui}
\IEEEauthorblockA{
\textit{Univ. Paris-Dauphine – Tunis} \\
Tunis, Tunisia \\
haroun.mezrioui@dauphine.tn}
\and
\IEEEauthorblockN{Yuyan Zhao}
\IEEEauthorblockA{
\textit{LAMSADE, Univ. Paris-Dauphine, PSL} \\
Paris, France \\
yuyan.zhao@dauphine.eu}
}

\maketitle

\begin{abstract}
Data provenance has numerous applications in the context of data preparation pipelines. It can be used for debugging faulty pipelines, interpreting results, verifying fairness, and identifying data quality  issues, which may affect the sources feeding the pipeline execution.
In this paper, we present an indexing mechanism to efficiently capture and query pipeline provenance.
Our solution leverages tensors to capture fine-grained provenance of data processing operations, using minimal memory. In addition to record-level lineage relationships, we provide finer granularity at the attribute level. This is achieved by augmenting tensors, which capture retrospective provenance, with prospective provenance information, drawing connections between input and output schemas of data processing operations. We demonstrate how these two types of provenance (retrospective and prospective) can be combined to answer a broad range of provenance queries efficiently, and show effectiveness through evaluation exercises using both real and synthetic data.
\end{abstract}


\section{Introduction}

Data preparation is widely recognized as a crucial phase in machine learning pipelines, as the effectiveness and quality of the resulting model often depend on how the training data is processed. In this context, provenance information can play an important role in tracing the lineage of datasets produced by the data preparation pipeline, facilitating understanding of data origins and aiding in debugging faulty pipelines. Accordingly, a number of proposals in the literature address the problem of provenance capture. Early work in this area focused on adding provenance support to scripting languages (see \cite{DBLP:journals/csur/PimentelFMB19} for a survey) to provide retrospective information on script execution, e.g., noWorkflow \cite{DBLP:conf/ipaw/MurtaBCKF14}, prospective provenance on the script specification, e.g., yesWorkflow \cite{549d902ea4984d6cb83950952c6ffaad}, or both types of information \cite{DBLP:conf/ipaw/PimentelDMBKMBL16}. These approaches primarily rely on execution log traces, which are gathered by the operating system as sequential events that capture file read and write actions, and/or through explicit annotations provided by the developer within the script to document main building blocks and relevant data dependencies.

While event logs collected by the operating system provide rich information, they are often low-level (e.g., file read and write events) and challenging to abstract and interpret within a data preparation pipeline. Additionally, these logs typically offer only coarse-grained provenance at the file level. In this regard, it is worth noting that adding file-grained provenance support for scripting languages, which involve numerous libraries and modules, can be challenging. Consequently, recent proposals have focused specifically on collecting provenance traces in data preparation pipelines, rather than in arbitrary scripts. These approaches target the modules and operations directly involved in data preparation, which often align closely with relational algebra operations, as seen in Python libraries like Pandas and Scikit-Learn. For example, Chapman {\em et al.} \cite{DBLP:journals/tods/ChapmanLMT24} recently released a library designed to capture fine-grained provenance (at the attribute-value level) for scripts performing data preparation tasks such as data transformation, fusion, augmentation, and reduction. Similarly, Grafberger {\em et al.} \cite{DBLP:journals/vldb/GrafbergerGSS22} proposed an approach that tracks provenance information for modules implemented using Scikit-Learn and explores the use of retrospective provenance to examine distributional changes between input and prepared datasets. Their goal is to identify skewness that may arise within data preparation pipelines, particularly concerning sensitive attributes like age or gender.

The above proposals assume and rely on persistent storage, as they target provenance applications where provenance is stored for long-term analysis and exploitation. In contrast, our work focuses on developing an in-memory solution that allows users to explore and utilize the provenance of the data preparation pipeline while it is still under development. Compared with disk-based solutions, an in-memory solution offers several advantages and is tailored for more immediate, dynamic use.  In particular, it allows users to explore and analyze the provenance of a data pipeline without waiting for its completion, enabling validation and issue identification during early stages of pipeline development. Additionally, it allows users to focus provenance capture on specific data processing steps that are more critical, such as those affecting data quality or fairness, thereby facilitating more targeted exploration and diagnosis. As such, it can facilitate the immediate detection and resolution of errors, such as incorrect transformations or anomalies in the data that might otherwise propagate through later stages of the pipeline if not addressed early. 
An in-memory solution also reduces latency in provenance queries by eliminating the overhead of disk-based read and write operations, resulting in faster response times when tracing the origins or transformations of specific data records.

Accordingly, the contributions of this research work are:

\begin{itemize}
    \item A tensor-based mechanism for capturing the fine-grained provenance of data preparation pipelines. Our solution records provenance at the level of individual data records (as opposed to complete files). In doing so, we adopt the ``why-provenance" flavor by tracking the origins and dependencies of each record. To enhance the efficiency of lineage query processing, we have optimized the representational structure of sparse binary tensors, enabling faster traversal of derivation relationships.

    \item Schematic annotations that specify the nature of data manipulations performed by each step in the data preparation pipeline. We use minimal representational means to capture not only the type of processing (and hence the how-provenance) but also the specific input and output columns affected in each dataset. This allows us to enhance the record-level provenance captured through tensors with additional attribute-level details, achieving finer-grained provenance, namely at the level of attribute-value, without the high overhead associated with explicitly tracking the provenance of each attribute value. 

    \item  Empirical evaluation exercises to assess the performance of our solution in terms of storage space required for encoding provenance, processing overhead for recording provenance, and processing provenance queries. Our evaluation uses three real-world data preparation pipelines and synthetic data from the TPC-DI benchmark\footnote{\url{https://www.tpc.org/tpcdi}}.

\end{itemize}

The paper is organized as follows.  We begin by
laying down the foundations of our work in Section \ref{sec:preliminaries}, where we introduce the data model used to capture key data processing operations in data preparation pipelines. In Section \ref{sec:tensor-provenance}, we present our solution, demonstrating how binary tensors can be utilized to capture the provenance of data processing steps in data preparation pipelines, and how schematic information at the level of columns can be integrated. We then specify how provenance queries can be processed using our solution in Section \ref{sec:querying}. Section \ref{sec:validatio} reports on the results of evaluation exercises that we conducted to assess the performance of our solution in terms of provenance capture and query processing. Finally, we review related work and compare it to ours in Section \ref{sec:relatedwork}, before closing the paper in Section \ref{sec:conclusions}.

\section{Foundations}
\label{sec:preliminaries}

Data preparation pipelines can be defined as a series of data processing steps. Depending on the nature of the processing performed, a processing step can take one or more datasets as input; however, it typically produces a single dataset. 
In our work, we consider popular data processing steps that are commonly used and referenced in the literature (see, e.g., \cite{GRL16,MWW19,DBLP:journals/tods/ChapmanLMT24}), and which can be categorized into seven categories: horizontal reduction, vertical reduction, horizontal augmentation, vertical augmentation, data transformation, join, and append. We will describe in what follows the semantics of each of these operations.

\begin{table}[h]
 \centering
 \vspace{-0.2cm}
 \caption{Main data preparation operations in ML pipelines.\\ (Table taken and adapted from \cite{DBLP:journals/tods/ChapmanLMT24})}
 \label{tab:taxonomy}
 \vspace{-0.3cm}
 \footnotesize
 \begin{tabular}{|c|c|}
  \hline
  {\bf Data Processing Operation}             & {\bf Category}                             \\ \hline
  Instance Selection                     & \multirow{3}{*}{Horizontal Data Reduction} \\ \cline{1-1} 
  Drop Rows                             &                                       \\ \cline{1-1} 
  Undersampling                         &                                       \\ \hline
  Feature Selection                     & \multirow{2}{*}{Vertical Data Reduction} \\ \cline{1-1} 
  Drop Columns                          &                                       \\ \hline
  Imputation                            & \multirow{5}{*}{Data Transformation}   \\ \cline{1-1} 
  Value transformation                  &                                       \\ \cline{1-1} 
  Binarization                          &                                       \\ \cline{1-1} 
  Normalization                         &                                       \\ \cline{1-1} 
  Discretization                        &                                       \\ \hline
  Instance Generation                   & \multirow{2}{*}{Horizontal Data Augmentation} \\ \cline{1-1} 
  Oversampling                          &                                       \\ \hline
  Space Transformation                  & \multirow{3}{*}{Vertical Data Augmentation} \\ \cline{1-1} 
  String Indexer                        &                                       \\ \cline{1-1} 
  One-Hot Encoder                       &                                       \\ \hline
  Join                                  & \multirow{2}{*}{Data Fusion}          \\ \cline{1-1} 
  Append                                &                                       \\ \hline
 \end{tabular}
 \vspace{-0.5cm}
\end{table}

\paragraph{Data Transformation} 
Operations in this category neither alter the schema of the dataset nor the number of records. Instead, they modify specific attribute values by applying a transformation function, e.g., binarization or normalization.

\paragraph{Vertical Reduction}
\texttt{feature selection} and \texttt{dropping columns} fall in this category. Both of these operations remove some of the attributes characterizing the data records in the input dataset $D^{in}$, giving rise to the output dataset $D^{out}$.

\paragraph{Vertical Data Augmentation}
Operations in this category include \texttt{Space Transformation}, \texttt{string indexer}, and \texttt{one hot encoder}. Applying vertical data augmentation to an input dataset $D^{in}$ produces a dataset $D^{out}$ with a different schema from $D^{in}$, but with the same number of records as $D^{in}$, with the $i^{th}$ record in $D^{out}$ corresponding to the $i^{th}$ record in $D^{in}$. 

\paragraph{Horizontal Reduction}
Given a dataset $D^{in}$, horizontal reduction produces a new dataset $D^{out}$, where the data records in $D^{out}$ are subsets of those in $D^{in}$: $D^{out} \subset D^{in}$. Data manipulations that fall into this category include \texttt{filter}, \texttt{instance selection}, and \texttt{undersampling}.

\paragraph{Horizontal Data Augmentation} Unlike the previous processing, horizontal augmentation adds new records. Examples of operations that fall into this category are \texttt{instance generation} and \texttt{oversampling}.

\paragraph{Join}
The \texttt{join} of the datasets $D^l$ and $D^r$, implemented using, e.g., the Merge operation in the Pandas library, and denoted by $D^l \Join^t_C D^r$, produces a dataset $D^j$ as a result of joining $D^l$ and $D^r$ on a boolean condition $C$, where $t$ represents the \texttt{join} type (inner, left outer, right outer, or full outer). Tables \ref{tab:dl},  \ref{tab:dr} and \ref{tab:d_j} illustrate a \texttt{join} example.

\begin{table}[h]
    \centering
    \vspace{-0.35cm}
    \begin{minipage}[t]{0.5\columnwidth} 
        \centering
        \caption{Dataset $D^l$}
        \label{tab:dl}
        \vspace{-0.3cm}
        \begin{tabular}{|c|c|c|c|}
            \hline
            & ID & Birthdate & Gender  \\ \hline
            1 & 10 & 1996-07-12 & F \\ \hline
            2 & 20 & 1994-03-08 & M  \\ \hline
            3 & 30 & $\bot$ & F  \\ \hline
            4 & 40 & 1987-11-23 & M  \\ \hline
        \end{tabular}
        \vspace{-0.2cm}
    \end{minipage}%
    \hfill 
    \begin{minipage}[t]{0.5\columnwidth} 
        \centering
        \caption{$D^r$ Dataset}
        \label{tab:dr}
        \vspace{-0.3cm}
        \begin{tabular}{|c|c|c|}
            \hline
            & ID & Name \\ \hline
            1 & 20 & Alice \\ \hline
            2 & 40 & Bob \\ \hline
        \end{tabular}
    \end{minipage}
\end{table}

\begin{table}[]
    \centering
    \vspace{-0.2cm}
    \caption{Join: $D^l \Join_{\text{inner}} D^r$}
    \label{tab:d_j}
    \vspace{-0.3cm}
    \begin{tabular}{|c|c|c|c|c|}
        \hline
        & ID & Birthdate & Gender  & Name \\ \hline
        1 & 20 & 1994-03-08 & M & Alice \\ \hline
        2 & 40 & 1987-11-23 & M & Bob \\ \hline
    \end{tabular}
    \vspace{-0.2cm}
\end{table}

\paragraph{Append}
The \texttt{append} operation, denoted by $D^l \uplus D^r$,
appends the records of a dataset $D^l$ at the end of the $D^r$. The two datasets do not need to have the same schema, and as such, the results are extended with null for the mismatching attributes. (This is similar to outer-union modulo ordering of the data records.) For example, appending $D^l$ to $D^r$ (in Tables \ref{tab:dl} and  \ref{tab:dr}) results in the dataset illustrated in Table \ref{tab:d_a}. 


\begin{table}[]
    \centering
     \vspace{-0.2cm}
    \caption{Append: $D^l \biguplus D^r$}
    \label{tab:d_a}
     \vspace{-0.3cm}
    \begin{tabular}{|c|c|c|c|c|c|}
        \hline
        & ID & Birthdate & Gender &  Name \\ \hline
        1 & 10 & 1996-07-12 & F  & $\bot$ \\ \hline
        2 & 20 & 1994-03-08 & M  & $\bot$ \\ \hline
        3 & 30 & $\bot$ & F  & $\bot$ \\ \hline
        4 & 40 & 1987-11-23 & M  & $\bot$ \\ \hline
        5 & 20 & $\bot$ & $\bot$  & Alice \\ \hline
        6 & 40 & $\bot$ & $\bot$  & Bob \\ \hline
    \end{tabular}
     \vspace{-0.5cm}
\end{table}

\section{Tensor-Based Indexing of Provenance}
\label{sec:tensor-provenance} 

In this section, we will begin by demonstrating how tensors can be utilized to encode provenance at the data record level, which corresponds to the \textit{why-provenance} flavors \cite{DBLP:journals/ftdb/CheneyCT09}. Next, we will illustrate how tensor representations can be adapted to efficiently address lineage queries. Finally, we will conclude the section by showing how tensors can be augmented with metadata to facilitate the inference of attribute-value provenance based on record-based provenance.
In the following, we assume that each data record $d$ is uniquely identified by the dataset it belongs to and its index within that dataset (typically corresponding to a row in a data frame). For instance, $D_i$ denotes the data record at index $i$ in the dataset $D$.

\subsection{Tensors for Recording Provenance}
To capture the provenance of data processing operations, we employ sparse tensors rather than full tensors, as most data processing operations involve only a single or a few output records derived from each input record. 
Tensors can be thought of as n-dimensional arrays. For our purposes, we use binary tensors. A binary tensor \(T\) of order \(p\) can be defined as a mapping from a multi-index \(K = K_1 \times \cdots \times K_p\) to \(\mathcal{B}\), where \(K_i \subset \mathcal{N}\), with \(\mathcal{N}\) denoting the set of positive natural numbers, and \(\mathcal{B}\) representing the set of boolean values: \(\mathcal{B} = \{0,1\}\).

\begin{definition}[Provenance of a Data Processing Operation]
\sloppy
Consider a data processing operation \(DP\) that consumed the input datasets \(D^{I_1}, \ldots, D^{I_n}\) and produced an output dataset \(D^O\), where \(n \geq 1\). The provenance of \(DP\) can be encoded using an order \(n + 1\) binary tensor \(T^{(DP)}\), such that \(T^{(DP)}(o,i_1, \ldots, i_n) = 1\) if the output record $D^{O}_{o}$  is provenance-wise derived from the input records indexed by \((D^{I_1}_{i_1}, \ldots, D^{I_n}_{i_n})\).
\end{definition}

The provenance of a given pipeline $P$ is represented using sets of binary tensors, with each tensor $T^{(DP)}$ associated with a data processing step  $DP$ in $P$. 
In the following sections, we examine the types of tensors generated for each class of data processing operations introduced in Section \ref{sec:preliminaries}.

\paragraph{Data Transformation} 
For a dataset \(D^{O}\) generated by transforming an input dataset \(D^{I}\), the provenance of this operation is represented by a 2-binary identity tensor.  This is because the input and output datasets \(D^{I}\) and \(D^{O}\) have the same shape (i.e., the same number of rows and columns), and the output record \(D^{O}_i\) corresponds provenance-wise to the input record indexed by the same value \(i\), i.e. \(D^{I}_i\), with $i \in [1,len(D^{I})]$.

\paragraph{Vertical Reduction}
Vertical reduction modifies the shape of the input dataset \(D^{I}\), resulting in an output dataset \(D^{O}\) with fewer columns. However, the number of records in \(D^{I}\) and \(D^{O}\) remains the same, with each output record \(D^{O}_i\) corresponding provenance-wise to the input record \(D^{I}_i\), where \(i \in [1, \text{len}(D^{I})]\). Therefore, similar to data transformations, vertical data reduction is represented using a 2-dimensional binary identity tensor.

\paragraph{Horizontal Reduction}
Horizontal data reduction modifies the shape of an input dataset \(D^{I}\), resulting in an output dataset \(D^{O}\) with fewer rows. Consequently, this transformation cannot be represented using a 2-dimensional binary identity tensor but instead requires a 2-dimensional binary masking tensor. Specifically, consider a 2-dimensional binary tensor \(T\) with dimensions \((|D^{O}| \times |D^{I}|)\) that encodes the provenance of the horizontal data reduction operation. In this tensor, some columns will contain only 0s, representing input records that were filtered out by the operation. Other columns, such as column \(j\), will contain a single cell \(T[i, j] = 1\), indicating that the input record \(D^{I}_j\) corresponds provenance-wise (and is identical in value in this case) to the output record \(D^{O}_i\).

\paragraph{Vertical Data Augmentation}
In vertical data augmentation, the input and output datasets, denoted by \( D^I \) and \( D^O \) respectively, have an equal number of rows, although \( D^O \) contains additional columns compared to \( D^I \). Each output record \( D^O_i \), where \( i \in [1, \text{len}(D^O)] \), is derived (in terms of provenance) from the data record with the same index in the input dataset, that is \( D^I_i \). Therefore, the provenance of vertical augmentation is captured using a 2-D binary identity tensor.

\paragraph{Horizontal Data Augmentation} 
In horizontal data augmentation, the output dataset \( D^O \) extends the input dataset \( D^I \) with additional records generated through oversampling or synthetic instance creation. For records in \( D^O \) that also appear in \( D^I \), provenance can be directly represented within the tensor. When new rows are produced by duplicating or slightly modifying existing instances (such as adding noise to numeric attributes while preserving key identifiers) their provenance can still be inferred with reasonable accuracy. Indeed, we strive in TensProv to maintain clear correspondences between the output records of the augmented dataset and the records of the input dataset, capturing provenance whenever such mapping can be explicitly established. 

\paragraph{Join}
The provenance of the \texttt{join} is captured as follows. Consider that $D^l$ and $D^r$ contain $n$ and $m$ records, respectively, and that dataset $D^j$ returned by the \texttt{join} has $p$ records. The provenance of the \texttt{join} is captured using a 3-binary tensor of size $(p \times n \times m)$. Specifically, $T[i,j,k] = 1$ if the \texttt{join} constructs the $i^{th}$ record in $D^j$ by combining the $j^{th}$ record in $D^l$ and the $k^{th}$ record in $D^r$.

As an example, consider the datasets $D^l$ and $D^r$ illustrated in Tables \ref{tab:dl} and  \ref{tab:dr}. The inner \texttt{join} of the two tables results in the dataset $D^j$ illustrated in Table \ref{tab:d_j}.
The provenance of the \texttt{join} can be represented by the following 3-binary tensor $T$ with dimensions $(2 \times 4 \times 2)$, in which the cells $T[2,1,1]$ and $T[4,2,2]$ take the value 1. Notice that the first dimension in the tensor refers to the data records in the results of the \texttt{join} ($D^j$), whereas the second and third dimensions refer to the input datasets $D^l$ and $D^r$. 
\[
T = 
\begin{pmatrix}
\begin{pmatrix}
0 & 0 \\
1 & 0 \\
0 & 0 \\
0 & 0
\end{pmatrix}, 
\begin{pmatrix}
0 & 0 \\
0 & 0 \\
0 & 0 \\
0 & 1
\end{pmatrix}
\end{pmatrix}
\]

\paragraph{Append}
The provenance of the \texttt{append} operation cannot be captured within a single tensor. Instead, it requires two tensors, one for each input dataset. Let $D^l$ and $D^r$ contain $n$ and $m$ records, respectively. The provenance is captured by two 2-binary tensors, each corresponding to one of the datasets. Such tensors can be thought of as a block diagonal tensor with zero padding. Specifically, the tensors capturing the provenance of \texttt{append} w.r.t. $D^r$ is of size $((n+m) \times n)$, and has a diagonal sub-tensor in the upper $(n \times n)$ block, while the lower block $(m \times n)$ block (from row $n+1$ to $n+m$) consists entirely of zeros. Conversely, the tensor capturing the provenance w.r.t. $D^l$ is of size $((n+m) \times m)$, with the lower $(m \times n)$ block (from row $n+1$ to $n+m$) being diagonal, while the upper block of size $(n \times m)$ is made up of zeros. 

\subsection{Provenance Capture}
Given a data processing operation $DP$, our goal is to construct a tensor (with the structure that we have just presented) to record its provenance. Broadly, two main approaches can be distinguished in doing so. The \emph{observation-based} approach derives provenance by examining and comparing the input and output datasets. In contrast, the \emph{active capture} approach modifies the input data by injecting annotations that are propagated through the operation, thereby producing provenance information directly.

Observation-based approaches are less intrusive but often computationally expensive. For instance, the provenance of a \texttt{join} can be reconstructed using hashing techniques to reduce computation costs: each input record is assigned a unique hash key, and each output record is decomposed into two parts according to the schemas of the input datasets. By matching hash keys, one can trace the originating records in the inputs. However, our experiments show that this approach remains computationally demanding for large datasets.

To address this limitation, we adopt a hybrid strategy that leverages the semantics of data processing operations. Specifically,
for operations like \texttt{filter}, which preserve data frame indices, we compare the indices of the records in the input and output datasets to reconstruct provenance efficiently. In contrast, for operations that do not preserve record indices, typically for the  \texttt{join} operation, we instrument the {\em active capture} method. We argue that this hybrid approach offers an efficient solution and a balanced trade-off, using active capture only when required while relying on operational semantics elsewhere to relate input and output records.

\subsection{Optimizing Tensor Representation for Querying}
\label{sec:tensor-indexing}

Sparse tensor representations, such as those provided by TensorFlow, aim to reduce space complexity by using two lists: one specifying the indices of non-zero cells in the tensor, and the second storing the actual values of these cells. Since we are dealing with binary tensors, the second list can be omitted entirely, leading to further memory savings.
That said, this type of representation is not well-suited for processing lineage queries, where the goal is typically to trace the index of a data record in one dataset to its corresponding indices in other datasets. In such cases, the query still needs to scan the list of indices in the tensor to identify the relevant record, which can be computationally expensive when the list is large.

\begin{figure}[tb]
\centering
\includegraphics[width=0.5\textwidth]{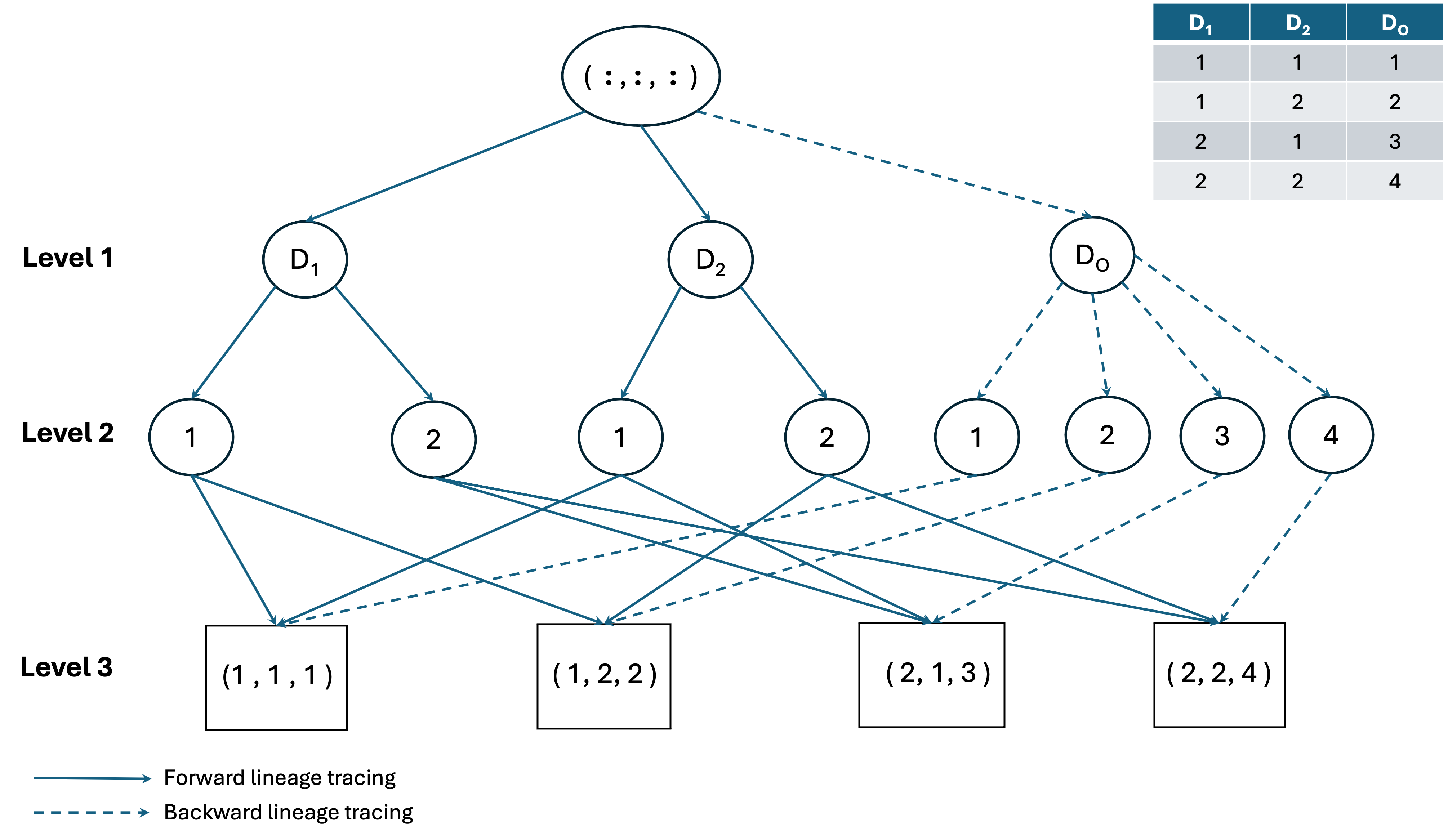}
\vspace{-0.6cm}
\caption{Tensor representation.}
\label{fig:tensor-tree}
\vspace{-0.5cm}
\end{figure}

To address the above limitation, we designed a new tensor representation that maintains memory usage close to that of existing sparse tensor formats (or potentially even less, as we omit the values list) while being optimized for efficient lineage query processing. Such a structure is depicted in Figure \ref{fig:tensor-tree}.

The tensor is used to represent the \texttt{join} of two datasets, $D_1$ and $D_2$, resulting in a new dataset $D_o$, as illustrated in the table at the top right of the figure. The values in the tensor correspond to indices in these datasets. For example, the first row indicates that record 1 in $D_o$ was produced by joining record 1 in $D_1$ with the record indexed at position 1 in $D_2$.
The tensor structure we adopt for such a \texttt{join} can be considered a special case of a tree, where leaf nodes can have multiple parents. More precisely, the tensor is represented using a rooted acyclic graph, with the root node representing the tensor. The children of the root (level 1 in the figure) represent the input and output datasets. The next level (level 2 in the figure) consists of the indices of the records bound to these datasets. Note that only the data records involved in the \texttt{join} are included at this level. Finally, the third and last level contains nodes that fully specify the indices of the input datasets that were combined in the \texttt{join}, as well as the index of the resulting output record.
Note that the number of nodes at this final level corresponds to the number of data records obtained as a result of the \texttt{join}, and these nodes are associated with multiple parents to minimize the memory required for encoding the provenance. 
Specifically, each node in the last level is associated with \(n + 1\) parents, where \(n\) represent the number of input  datasets involved in the corresponding data processing operation. Solid edges in the graph represent links that are followed for evaluating forward lineage queries, whereas dashed edges represent the edges used for backward lineage tracing.

Let us now examine the complexity in terms of memory as well as in terms of lineage tracing using the structure we have just presented.
The memory complexity is $O(T \times d)$, where $T$ represents the number of true values and $d$ is the number of dimensions. This is also the space complexity for sparse tensor structures, such as the one provided by TensorFlow. 
However, using our structure, lineage queries can be performed quickly with just three list accesses, corresponding to the height of the tensor's tree-like structure. In contrast, querying with TensorFlow's sparse tensor representation is  more expensive, as it requires scanning the entire index array, resulting in a time complexity of $O(T)$. We can estimate the size of $T$ more precisely based on the semantics of the data processing operation. For instance, in the case of a \texttt{join}, $T$ corresponds to the size of the dataset produced by the \texttt{join} operation: each record in the output dataset is associated with a pair of records from the input datasets. Hence, in this case, $T = |D_o|$.

\paragraph{From Bag-Semantics to Set-Semantics}
It is worth noting that the tensor representation we have just presented captures the \texttt{join} as implemented by the `merge` operation of Pandas, which is the most popular method used in Python for joining datasets and is known to follow the bag-semantics of joins. This means that two identical records produced by different pairings of input data records are preserved separately and associated with different indices in the output dataset. However, in some cases, to provide users with better lineage insight, a set-based processing of provenance queries might be required. 

To illustrate this, consider the \texttt{join} depicted in Figure \ref{fig:tensor-tree}, where the data records indexed by $2$ and $4$ are identical. In such a case, when a user queries for the input data records responsible for the creation of the data record indexed by $2$, or more specifically, the data record itself (not its index), we should return not only the input data records that led to the creation of the data record identified by index $2$, but also those responsible for the creation of the data record indexed by $4$.

To meet the set-semantics, 
the tensor representation can be modified accordingly. Specifically, for duplicate records in the output dataset, we would use a single ID, such as the smallest one. For example, in the case illustrated in Figure \ref{fig:tensor-tree}, the ID $4$ would be replaced by $2$ (assuming the two data records are duplicates). This adjustment reduces the number of nodes at level $2$, specifically those associated with the output dataset. In our example, instead of having four values associated with dataset $D_o$, we would have three, as the duplicate records indexed by $2$ and $4$ would both be represented by index $2$.

\subsection{Attribute-Based Provenance Capture}

Using tensors, as presented in the previous section, allows  capturing the provenance, or more specifically the derivation relationships, between the data records of datasets that are used and generated by preprocessing data manipulations. In this section, we show how tensors can be augmented with metadata information to cater to finer-grained provenance capture at the level of attribute values.

\paragraph{Horizontal Data reduction} Tensors are sufficient for inferring attribute-based derivations for data manipulations of this type. Indeed, for both \texttt{feature selection} and \texttt{subsampling}, the data records in the output dataset remain identical in value to those in the input datasets. Consequently, the derivation relationships can be reliably inferred between attribute values that occupy the same positions in  the output data record and its corresponding input data record.

\paragraph{Vertical Data reduction} For data manipulations of this kind, unlike the previous case, we require metadata to infer attribute-based provenance. This can be achieved by identifying the attributes retained by the reduction operation. A bitset can be used for this purpose, where its size corresponds to the number of features in the input dataset, and each bit is set to $1$ if the corresponding attribute is preserved by the reduction. For example, the bitset $10011$ indicates that out of five features in the input dataset, the reduction operation retained the 1st, 4th, and 5th attributes. 

It is important to note that the above assumes that the order of attributes in the input and output datasets is preserved, which is generally the case in vertical reduction manipulations. If this assumption does not hold, a list of integers can be used instead of a bitset, specifying the attributes retained from the input and the order in which they appear in the output dataset. For example, [4, 2, 5] indicates that three attributes were retained, and the order in the output dataset was altered so that the 4th attribute appears before the 2nd, which appears before the 5th attribute of the input dataset.

\paragraph{Vertical Augmentation} 
Operations that fall under this category, such as \texttt{space transformation}, \texttt{string indexer}, and \texttt{one hot encoder}, add new attributes \(Y\) by applying a function to existing attributes \(X\) within the input data record: \(Y = f(X)\). To support attribute-based provenance, it is essential to provide information about the newly added attributes \(Y\). This can be represented using a bitset. For example, a bitset like \(000011\) indicates that the input dataset contains four attributes, and the fifth and sixth attributes were added through vertical augmentation. 
Additionally, it is important to specify which input attributes were used to compute the values of the new attributes. This information can also be represented with a bitset. For instance, the bitset \(1010\) indicates that the values for attributes \(Y\) were computed using the first and third attributes of the input dataset. This information is sufficient to establish attribute-value correspondences between the attributes present in both the input and output datasets, as well as the newly added attributes. 

In fact, we do not need to maintain two separate bitsets: one for the input attributes and another for the output attributes. Instead, we can use a single bitset \(101011\), which, combined with the knowledge that the input dataset has four attributes, indicates that the first and third attributes of the input dataset were used to generate the two new attributes.

\paragraph{Data Transformation}
The shapes of the input and output datasets are identical. As previously discussed, in this scenario, a data record \(D^O_i\) in the output dataset \(D^O\) is lineage-wise dependent on the corresponding data record \(D^I_i\) in the input dataset \(D^I\) with the same index. Furthermore, for an attribute located at position \(j\), the value \(D^O_i[j]\) is lineage-wise derived from the attribute value at the same position in the corresponding input record, \(D^I_i[j]\). Consequently, in the case of data transformation operations, no additional metadata is required to ascertain the provenance at the attribute value level.

\paragraph{Horizontal Augmentation}
As with data transformation, we assume that each attribute in the output dataset corresponds, schema-wise, to the attribute in the same position in the input dataset.

\paragraph{Join}
For the \texttt{join} operation, we only need a method to specify the correspondences between the columns (features) of the output dataset and those of the input datasets. Consider, for instance, that $D^o$ was obtained by joining $D^l$ with $D^r$. To determine attribute-based provenance, we only require two bitsets, one for each input dataset. The size of each bitset corresponds to the number of attributes in the output dataset. For example, if $D^o$ contains six attributes, a list of bitsets of size 2 is associated with the tensor. The list $[10101, 11010]$, for instance, indicates that the first attribute in the output dataset $D^o$ corresponds to the first attribute in both $D^l$ and $D^r$, the second attribute in $D^o$ corresponds to the second attribute in $D^r$, and so on.

\paragraph{Append}
Consider that the dataset $D^o$ was obtained by appending $D^r$ to $D^l$. Similar to the \texttt{join} operation, to infer the attribute-based provenance, we require two bitsets that specify the correspondences between the attributes of $D^o$ and those of $D^l$ and $D^r$. 

Based on the discussion above, it becomes clear that certain data processing operations require additional metadata to derive attribute-value provenance from record provenance. These operations include vertical data reduction, vertical data augmentation, and \texttt{join} operations (we omit the \texttt{append} operation since its treatment is similar to the case of the \texttt{join}). Table \ref{tab:attribute-metadata} outlines the specific metadata that needs to be added for each of these operations.

\begin{table}[ht]
\footnotesize
\vspace{-0.2cm}
\caption{Metadata enabling the inference of attribute-value provenance from record-based provenance.}
\label{tab:attribute-metadata}
\vspace{-0.25cm}
\begin{tabularx}{\columnwidth}{|c|>{\raggedright\arraybackslash}X|>{\raggedright\arraybackslash}X|} 
\toprule
\textbf{Operation} & \textbf{Bitset} & \textbf{Comment} \\ 
\midrule
\textbf{Vert. reduct.} & 10011 & The 2nd and 3rd attributes are dropped. \\ 
\midrule
\textbf{Vert. augment.} & \textcolor{blue}{1010}\hspace{0pt}\textcolor{purple}{11} & Knowing that the input dataset has 4 attributes, we deduce that the 1st and 3rd attributes were used to engineer two additional attributes. \\ 
\midrule
\multirow{2}{*}{\textbf{Join}} & 10101 & The 1st, 3rd, and 5th attributes in the output dataset come from the 1st input dataset. \\ 
\cline{2-3}
& 11010 & The 1st, 2nd, and 4th attributes in the output dataset come from the 2nd input dataset. \\ 
\bottomrule
\end{tabularx}
\vspace{-0.5cm}
\end{table}

\subsection{Localized vs. Contextual Data Processing Operations}
\label{sec:localvscontextual}
In our solution, we aim to minimize the storage of datasets used and generated by the pipeline operations, retaining them only when necessary. Specifically, we store, in the majority of cases, only the datasets that serve as input and output for the entire pipeline, and only occasionally some intermediate datasets, as we shall explain later. This approach is feasible because the provenance captured by the tensors allows intermediate results to be recomputed when needed. That said, to enable efficient recomputation of the intermediate data record at querying time, we distinguish between two types of data processing operations: \textbf{localized} and \textbf{contextual}.

    \textbf{Localized Operations}: In these operations, an output data record can be recomputed by applying the operation to its provenance-related input data records (which are identified by the tensor associated with the operation). For example, vertical data reduction operations fall into this category. Since the input data records required for recomputation are explicitly defined by the tensor, there is no need to materialize the input datasets for localized operations.

    \textbf{Contextual Operations}: These operations, such as \texttt{data imputation}, require information from the entire dataset to compute an output data record. For instance, when imputing missing values using the mean or median, the computation depends on the values of the attribute across the entire input dataset, not just the provenance-related input records specified by the tensor. For contextual operations, we materialize their input datasets to enable efficient recomputation at query time.

By materializing the input datasets of contextual operations, we strike a balance between: \textbf{i) Storing all datasets} used and generated by the pipeline steps, which is memory intensive and represents a limitation of state-of-the-art solutions, and \textbf{ii) Relying solely on tensors} for provenance indexing, which, at query time, could require re-executing the entire pipeline on the full input datasets. Indeed, our solution ensures efficient query processing by allowing recomputation to be performed only on the relevant input data records (for backward lineage queries) or output data records (for forward lineage queries), rather than reprocessing the entire dataset.

\section{Querying Provenance Information}
\label{sec:querying}

\begin{figure}[tb]
\centering
\includegraphics[width=0.47\textwidth]{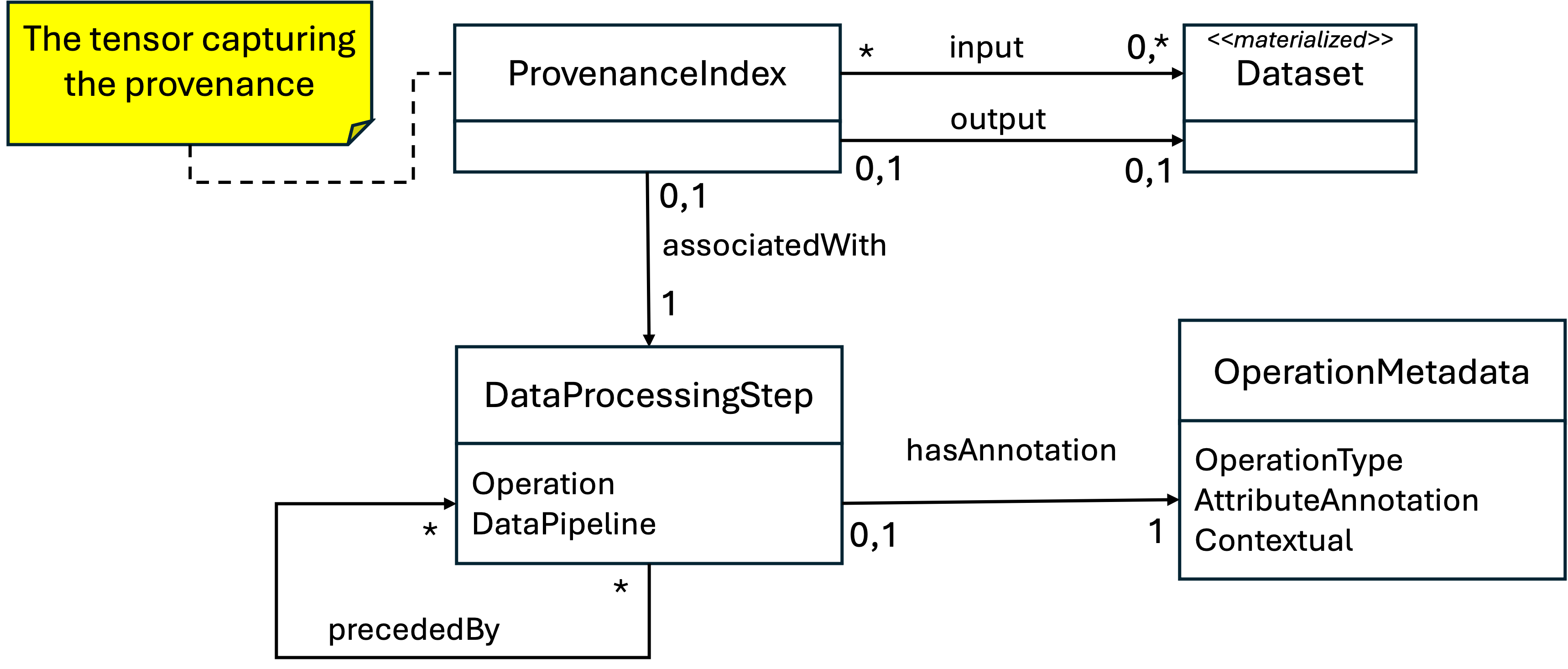}
\vspace{-0.3cm}
\caption{Elements of the solution put together.}
\label{fig:model}
\vspace{-0.5cm}
\end{figure}

Before presenting the provenance queries to be examined in the context of data preparation pipelines, it is useful to first consolidate the concepts discussed so far within a unified model, as illustrated in Figure~\ref{fig:model}. The class \texttt{ProvenanceIndex} represents the tensor used to capture provenance. This tensor is associated with information about the data processing steps from which it captures provenance, as well as the data pipeline to which each operation belongs. 
Note that the data processing steps are related to each other through precedence relationships, which reflect their dependencies within the data pipeline. The processing steps are further enriched with metadata specifying the type of data processing operation, such as vertical augmentation, horizontal reduction, and data transformation, and specifying if the operation is contextual. They are also associated with attribute annotations in the form of bitsets when the operation in question performs vertical data reduction, vertical data augmentation, or a join, as discussed in the previous section.
The provenance index also references the datasets used as inputs and produced as outputs by these operations, when such datasets are materialized. Indeed, as detailed earlier, only datasets that are used as input to \texttt{contextual} operations are materialized.

Now, regarding provenance queries, we seek a mechanism to trace the lineage of an input-output data item both backward and forward within a pipeline execution. Various queries have been explored in the literature for this purpose. Such queries can be categorized into the following classes:

\begin{itemize}[leftmargin=*, itemsep=0pt, topsep=0pt]
\item
\textbf{Forward Lineage Tracing}:  
Consider two datasets, \( D^1 \) and \( D^2 \), where \( D^2 \) is derived from \( D^1 \) through the dataflow of a data preparation pipeline. Given a data record \( d \) in \( D^1 \), the objective is to identify the records in \( D^2 \) to which \( d \) contributed.

\item
{\bf Backward lineage tracing}: 
Consider two datasets, \( D^1 \) and \( D^2 \), where \( D^2 \) is derived from \( D^1 \) through the dataflow of a data preparation pipeline. Given a data record \( d \) in \( D^2 \), the objective is to identify the records in \( D^1 \) that contributed to the existence of \( d \).

\item
\textbf{Co-contributory Lineage Tracing}:  
Consider three datasets, \( D^1 \), \( D^2 \), and \( D^3 \), where elements in \( D^3 \) are generated from combinations of elements in \( D^1 \) and \( D^2 \). Given a data record \( d \) in \( D^1 \), the objective is to identify the specific records in \( D^2 \) that were used in conjunction with \( d \) to produce corresponding records in \( D^3 \).

\item

\textbf{Co-dependency Lineage Tracing}:  
Consider three datasets, \( D^1 \), \( D^2 \), and \( D^3 \), where the elements in \( D^2 \) and \( D^3 \) have lineage dependencies, i.e. generated from, records in \( D^1 \). Given a data record \( d \) in \( D^2 \), the objective is to trace and identify the records in \( D^3 \) that are lineage-dependent on those records in \( D^1 \) which contributed to \( d \).

\end{itemize} 

The above classes of queries are defined at the level of data records. A finer-grained version can be defined at the attribute-value level, providing more detailed insights. Additionally, these queries can be enhanced with information on the operations used during processing, thereby capturing both the \textit{how}-provenance (the process details) in addition to the \textit{why}-provenance (the source records/attribute-values). Table~\ref{tab:provenance_queries} lists the classes of provenance queries we consider, and that are relevant in the context of data preparation pipelines.

\begin{table*}[]
\footnotesize
\caption{Provenance queries.}
\label{tab:provenance_queries}
\vspace{-0.35cm}
\begin{tabularx}{\textwidth}{|c|X|c|c|X|}
\hline
\textbf{ID} & \multicolumn{1}{c|}{\textbf{Provenance Query}} & \multicolumn{1}{c|}{\textbf{Granularity}} & \multicolumn{1}{c|}{\textbf{Input}} & \multicolumn{1}{c|}{\textbf{Output}} \\ \hline
\textbf{Q1} & Why-Provenance (forward) & Data record & $D_i$ & What are the data records that were derived from $D_i$ \\ \hline
\textbf{Q2} & Why-Provenance (backward) & Data record & $D_i$ & What are the data records that contributed to $D_i$ \\ \hline
\textbf{Q3} & Why-Provenance (forward) & Attribute-value & $D_i[att_j]$ & What are the attribute values derived from $D_i[att_j]$ \\ \hline
\textbf{Q4} & Why-Provenance (backward) & Attribute-value & $D_i[att_j]$ & What are the attribute values that contributed to $D_i[att_j]$ \\ \hline
\textbf{Q5} & How-Provenance (forward) & Data record & $D_i$ & What are the data records derived from $D_i$ with the associated operations \\ \hline
\textbf{Q6} & How-Provenance (backward) & Data record & $D_i$ & What are the data records that contributed to $D_i$ with the associated operations \\ \hline
\textbf{Q7} & How-Provenance (forward) & Attribute-value & $D_i[att_j]$ & What are the attribute values derived from $D_i[att_j]$ with the associated operations \\ \hline
\textbf{Q8} & How-Provenance (backward) & Attribute-value & $D_i[att_j]$ & What are the attribute values that contributed to $D_i[att_j]$ with the associated operations \\ \hline
\textbf{Q9} & All transformations & Dataset & $D$ & What are the operations applied to the dataset $D$ \\ \hline
\textbf{Q10} & Co-contributory lineage & Data record & $D^1_i, D^2$ & What are the records in $D^2$ used with $D^1_i$ to create new records \\ \hline
\textbf{Q11} & Co-dependency lineage & Data record & $D^1_i, D^2$ & What are the records in $D^2$ that can be traced back to data records that were responsible for the generation of $D^1_i$ \\ \hline
\end{tabularx}
\vspace{-0.4cm}
\end{table*}

We do not rely on disk storage to store and query our provenance. Instead, we directly process queries in memory by mainly leveraging the tensors that index the provenance, and in some cases, also on the metadata associated with the tensors for certain types of queries. For example, in data-record-level queries (i.e., $Q_1$, $Q_2$, $Q_5$, and $Q_6$), the tensors alone are used by following their tree structure and leveraging the relationships between operations (the precedence relationships shown in Figure \ref{fig:model}) to navigate between tensors until reaching the datasets of interest.

For attribute-level queries (i.e., $Q_3$, $Q_4$, $Q_7$, and $Q_8$), the metadata associated with operations is additionally utilized to drill down from data records to attribute values. Query $Q_9$ relies mainly on the relationships between operations, and thus does not involve the specifics of each tensor in detail.

Finally, co-contributory and c-influence lineage queries (i.e., $Q_{10}$ and $Q_{11}$) employ a combination of forward processing as in $Q_1$ and backward processing as in $Q_2$.

To evaluate provenance queries, we make use of two main operations on tensors, namely slicing and projection. 

\paragraph*{Tensor Projection}
Let \(\text{T} \in \{0,1\}^{I_1 \times I_2 \times \dots \times I_N}\) be an \(N\)-order binary tensor with dimensions \(I_1, I_2, \ldots, I_N\). \footnote{While most tensors  will be 2D or 3D, our method is not limited to these dimensions. It can also handle manipulations involving more than three datasets, e.g., the provenance of a three-way join involving three input datasets and producing one dataset would be represented by a 4D binary tensor.} The projection of \(\text{T}\) on a subset of dimensions \(\{d_1, d_2, \ldots, d_m\}\), where \(m < N\), results in a new binary tensor \(\text{P} \in \{0,1\}^{I_{d_1} \times I_{d_2} \times \dots \times I_{d_m}}\). Each element \(p_{i_{d_1} i_{d_2} \dots i_{d_m}}\) of \(\text{P}\) is obtained by applying a logical OR over elements along the remaining \(N - m\) dimensions of \(\text{T}\):\vspace{-0.3cm}
\[
p_{i_{d_1} i_{d_2} \dots i_{d_m}} = \bigvee_{\substack{i_n = 1 \\ n \notin \{d_1, d_2, \dots, d_m\}}}^{I_n} t_{i_1 i_2 \dots i_N}
\]
\vspace{-0cm}
for \(i_{d_1} \in \{1, \ldots, I_{d_1}\}, \ldots, i_{d_m} \in \{1, \ldots, I_{d_m}\}\).
Here, the logical OR (\(\vee\)) operation ensures that the resulting projected tensor \(\text{P}\) remains binary.
In what follows, we denote the projection of a tensor \(\text{T}\) along a dimension \(p\) by \(\text{project}(\text{T}, p)\).

\paragraph*{Tensor Slicing}
Slicing a tensor involves selecting a subset of its elements by fixing one or more dimensions to specific values. Given a binary tensor \(\text{T} \in \{0,1\}^{I_1 \times I_2 \times \dots \times I_N}\) of order \(N\), a slice can be obtained by fixing certain indices. For example, slicing \(\text{T}\) along the first dimension at \(i_1 = i_0\) results in an \((I_2 \times I_3 \times \dots \times I_N)\)-dimensional subtensor \(\text{T}_{i_0}\): \vspace{-0.2cm}
\[
\text{T}_{i_0} = \left[ t_{i_0 i_2 i_3 \dots i_N} \mid i_2 \in \{1, \ldots, I_2\}, \dots, i_N \in \{1, \ldots, I_N\} \right]
\]
In our work, we will slice the tensor by fixing a given dimension for one or multiple indices \(i_{p1}, i_{p2}, \ldots, i_{pm}\), which gives a lower-order subtensor: 
\[
\begin{split}
\text{T}_{\{i_{p1}, i_{p2}, \ldots, i_{pm}\}} 
    &= \left[ t_{i_1 i_2 \dots i_N} \mid i_p \in \{i_{p1}, i_{p2}, \ldots, i_{pm}\}, \right. \\
    &\qquad \left. \forall i_j \in \{1, \ldots, I_j\} \text{ for } j \neq p \right]
\end{split}
\]
Let \(S = \{i_{p1}, i_{p2}, \ldots, i_{pm}\}\). We will denote the above operation by \(\text{slice}(\text{T}, p, S)\), where \(p\) denotes the dimension along which the tensor is sliced. 

For exposition's sake, we slightly abuse notation and write 
\(\text{slice}(\text{T}, p, d_s)\) (resp. \(\text{slice}(\text{T}, p, D_s)\)) 
to denote the tensor obtained by fixing dimension \(p\) of \(\text{T}\) 
to the index (resp. indices) corresponding to the data record \(d_s\) 
(resp. records in \(D_s\)).

\subsection*{Processing Record-Based Provenance Queries}

In what follows, we focus on the processing of query $Q_1$ (other data-record-based queries are handled in a similar manner).  
Let $D$ be a dataset and $D'$ another dataset such that $D'$ is reachable from $D$ through a data pipeline.  
Let $d_s$ be a data record in $D$. To determine which records in $D'$ are derived from $d_s$, we examine the dataflow path connecting $D$ to $D'$.  
This dataflow consists of a sequence of data-dependent operations  
$(Op_1, \ldots, Op_n)$, 
executed in order, where $Op_1$ takes $D$ as input and $Op_n$ produces $D'$ as output.  
Each operation $Op_j$ is associated with a tensor $T_j$ that encodes its provenance. The overall dataflow can thus be represented as the following  sequence: \vspace{-0.15cm} 
\[
\bigl( (T_1, p_i^{T_1}, p_o^{T_1}), \ldots, (T_n, p_i^{T_n}, p_o^{T_n}) \bigr),
\]  
where $T_j$ ($j \in [1,n]$) denotes the provenance tensor corresponding to operation $Op_j$, $p_i^{T_j}$ represents a dimension associated with its inputs, and $p_o^{T_j}$ a dimension associated with its outputs.  
Specifically, the dataset $D$ corresponds to the input dimension $p_i^{T_1}$ of $T_1$, while $D'$ corresponds to the output dimension $p_o^{T_n}$ of $T_n$.
To identify the data records in $D'$ that are derived from a specific record $d_s$ in $D$, it is sufficient to examine the tensors along the data flow path $((T_1, p_i^{T_1}, p_o^{T_1}), \ldots, (T_n, p_i^{T_n}, p_o^{T_n}))$ as follows.
Starting with the data record $d_s$, we examine the first tensor $T_1$ to identify the records associated with the dataset in dimension $p_o^{T_1}$. This is achieved by performing a slice operation on $T_1$ along dimension $p_i^{T_1}$, retaining only the element associated with $d_s$. Then, we apply a projection to isolate the data records, specifically the identifiers of records associated with dimension $p_o^{T_1}$. Formally, this operation can be expressed as:
$D_t = \text{project}(\text{slice}(T_1, p_i^{T_1}, d_s), p_o^{T_1}).$
The data records obtained from this step are then used as input for the next tensor in the path, $T_2$, as follows:
$\text{project}(\text{slice}(T_2, p_i^{T_2}, D_t), p_o^{T_2})$.
This process is repeated for each tensor along the path until reaching the final tensor $T_n$. At this point, the projection will yield the data records in $D'$ that are derived from the original record $d_s$.

Note that the tensor structure introduced in Section~\ref{sec:tensor-indexing} enables efficient processing of slicing and projection operations. 
Consider, for instance, a join tensor \( T \) that represents the join between two datasets \( D^l \) and $D^r$, producing a resulting dataset \( D^o \). 
Suppose we are given a record \( d_s \in D^l \) and we wish to identify the corresponding records in \( D^o \). 
This can be achieved in two main steps. 
First, we slice the tensor \( T \) to retain only the cells that refer to \( d_s \). 
This operation is performed by traversing the tensor structure illustrated in Figure~\ref{fig:tensor-tree} from the root to the leaves, which involves three access levels. 
Next, project the sliced tensor to keep only the components corresponding to \( D^o \). 
This projection is straightforward: we simply extract the third (i.e., last) position of the triples contained in the previously selected cells.

In the above method for processing provenance queries, we focused on the provenance of a single data record, which can be generalized in a straightforward manner to a specific set of records. In some applications, however, users may need to identify, for the entire output dataset $D'$ of a pipeline, the corresponding records in its input dataset $D$. This is useful, e.g., when assessing fairness (e.g., determining the proportion of female/male individuals in the output dataset using a gender attribute available only in the input dataset),  or for making sure that the output records originate from input records of individuals who provided consent. For such cases, instead of performing successive slicing and projection as shown earlier, we employ Einstein summation \cite{da2025tensors} between the tensors representing the provenance of the operations within the data preparation pipeline. \vspace{-0.15cm} 
\[
T_1 \otimes_{p_o^{T_1} = p_i^{T_2}} 
    T_2 \otimes_{p_o^{T_2} = p_i^{T_3}} 
    \cdots 
    \otimes_{p_o^{T_{n-1}} = p_i^{T_n}} 
    T_n
\] 
\vspace{-0.05cm}
\noindent
This yields a tensor capturing the correspondences between the input dataset $D$ and the output dataset of $D'$.

\subsection*{Processing Attribute-Based Provenance Queries}
In this section, we show that by using bitsets associated with the data processing operations (introduced earlier in Table~\ref{tab:attribute-metadata}), we can define mappings between the attributes of input and output datasets. This allows refining the granularity of provenance queries from the data-record level down to individual attribute-value cells.

Here, we note that some data processing operations (specifically, horizontal reduction, horizontal augmentation, data transformations, and data append) preserve the positional mapping of attributes between input and output datasets. Specifically, given an output data record \(d'\) derived from an input data record \(d\) using any of these operations, the attribute at position \(i\) in \(d'\), denoted \(d'[i]\), is derived directly from the attribute at position \(i\) in \(d\), i.e., \(d[i]\). Thus, the mapping function for attributes between the input and output of these operations is effectively the identity function. As such, these kinds of operations are not associated with bitsets.

In contrast, for the remaining data processing operations, in particular vertical reduction, vertical augmentation, and join, the attribute mapping functions between input and output records are not necessarily identity mappings\footnote{We do not specify the mapping for the \texttt{append} operation, as it is similar to the mapping used in the case of vertical augmentation, applied separately to each of the two datasets involved in the \texttt{append}.}. In what follows, we detail these mappings, which rely on the bitset scheme introduced in Table~\ref{tab:attribute-metadata}. For each type of data manipulation, we distinguish \textit{forward mappings}, linking input dataset attributes to their counterparts in the output dataset, and \textit{backward mappings}, linking output attributes back to their source attributes in the input dataset.

\subsubsection*{Mapping Attributes in Vertical Reduction}
As noted earlier in Table~\ref{tab:attribute-metadata}, a vertical reduction operation is annotated by the bitset \(\mathbf{b} = (b_1, b_2, \dots, b_n)\), where each \(b_i \in \{0,1\}\) denotes whether the \(i\)-th attribute is included (\(1\)) or excluded (\(0\)) in the reduction.

\paragraph{Forward Mapping}
The function \(\text{map}^{\text{f}}_{\text{vr}}(\mathbf{b}, i)\) takes as input the bitset \(\mathbf{b}\) associated with the vertical partitioning operation and the position \(i\) of an attribute in the input schema. It returns the position of the corresponding attribute in the output schema as follows:
\vspace{-0.25cm}
\[
\text{map}^{\text{f}}_{\text{vr}}(\mathbf{b}, i) =
\begin{cases}
\text{null}, & \text{if } b_i = 0, \\
\sum_{k=1}^{i} b_k, & \text{if } b_i = 1.
\end{cases}
\]
\noindent
That is, it returns \(\text{null}\) when \(b_i = 0\) (meaning the attribute was removed as part of the vertical reduction); otherwise, it yields the number of ones in \(\mathbf{b}\) up to and including position \(i\), i.e., the count of set bits preceding \(i\) plus one.

\paragraph{Backward Mapping}
The function \(\text{map}^{\text{b}}_{\text{vr}}(\mathbf{b}, i)\) takes as input the bitset \(\mathbf{b}\) and an index \(i\), representing the \(i\)-th attribute in the output dataset. It returns the position \(j\) of the corresponding attribute in the input dataset, where \(j\) satisfies the following conditions:\vspace{-0.25cm}
\[
\sum_{k=1}^{j} b_k = i \quad \text{and} \quad b_j = 1.
\]
\vspace{-0.25cm}
\subsubsection*{Mapping Attributes in Vertical Augmentation}

Let the bitset \(\mathbf{b} = (b_1, b_2, \dots, b_n)\) annotate a vertical augmentation operation (see Table~\ref{tab:attribute-metadata}), and assume the input dataset has \(m\) attributes, with \(m < n\).

\paragraph{Forward Mapping}  
For vertical augmentation, the forward mapping is the identity function. Indeed, using vertical augmentation, all attributes of the input dataset are preserved, retaining the same positions in the output dataset.

\paragraph{Backward Mapping}  
The function \(\text{map}^{\text{b}}_{\text{va}}(\mathbf{b}, m, i)\) takes as input the bitset \(\mathbf{b}\), the number of attributes \(m\) in the input dataset, and a natural number \(i\) representing the position of an attribute in the output dataset.  
If \(i \leq m\), the corresponding attribute in the input dataset has the same position \(i\).  
Otherwise, when \(i > m\), the output attribute is newly derived from one or more input attributes. Specifically, such input attributes are associated with the bits that are set in the bitset \(\mathbf{b}\), that is: \vspace{-0.25cm}
\[
\{\, j \mid (j \leq m) \wedge (b_j=1) \,\}.
\] 

\vspace{-0.15cm}

\sloppy
\subsubsection*{Mapping of Attributes in Join Operation}

Let the bitset \(\mathbf{b} = (b_1, b_2, \dots, b_n)\) annotate a \texttt{join} operation, or more specifically, the way the attributes of an input dataset appear in the results of the \texttt{join}.

\paragraph{Forward Mapping}  
The forward mapping function $\text{map}^{\text{f}}_{\text{join}}(\mathbf{b}, i)$ takes as input the bitset $\mathbf{b}$ and a natural number $i$, representing the position of an input attribute. It returns an integer $j$, which denotes the position of the corresponding output attribute, and which  satisfies the following conditions:\vspace{-0.3cm} \[
\sum_{k=1}^{j} b_k = i \quad \text{and} \quad b_j = 1.
\]
\vspace{-0.2cm}

\paragraph{Backward Mapping}  
The backward mapping function $\text{map}^{\text{b}}_{\text{join}}(\mathbf{b}, i)$ takes as input a bitset $\mathbf{b}$ and a position $i$ of an attribute in the output dataset. It returns the corresponding position of that attribute in the input dataset, as follows:
\vspace{-0.1cm}
\[
\text{map}^{\text{b}}_{\text{join}}(\mathbf{b}, i) = 
\begin{cases}
\text{null}, & \text{if } b_i = 0, \\
\sum_{k=1}^{i} b_k, & \text{if } b_i = 1.
\end{cases}
\]
\vspace{-0.05cm}
If $b_i = 0$, the function returns $\text{null}$, indicating that the $i$-th output attribute does not originate from this input dataset in question. If $b_i = 1$, it returns the number of set bits (ones) from position $1$ to $i$, effectively giving the rank of the attribute among the included attributes in the \texttt{join}.

\section{Validation}
\label{sec:validatio}

There are mainly two approaches for implementing provenance capture.  
In the first approach, the user explicitly calls a \texttt{capture} method whenever a new \texttt{DataFrame} is derived from an existing one.  
This manual tracking strategy tends to clutter the user’s code and introduces some overhead from the developer’s perspective.  
To overcome this problem, some provenance capture systems attempt to make the tracking automatic, for example, by overriding data manipulation methods so that provenance is captured in addition to performing the data processing itself.  
This is, for instance, the approach adopted by the MLInspect system~\cite{DBLP:journals/vldb/GrafbergerGSS22}.  
However, our experiments with MLInspect revealed that this approach struggles to remain functional over time as Python libraries evolve; e.g., we were unable to run MLInspect with Python version~3.13.2.  

In our work, we adopted a different strategy based on the \textit{decorator design pattern}.  
We wrap a \texttt{pandas.DataFrame} object in a container class that proxies all method calls and attribute accesses.  
This allows our system to intercept operations that create or modify \texttt{DataFrames} and to automatically capture their provenance.  
A similar idea is employed by Chapman \textit{et al.}~\cite{DBLP:journals/tods/ChapmanLMT24}, but our implementation differs in the follwing aspects.  
In Chapman \textit{et al.}’s approach, the \texttt{DataFrame} is captured both before and after each manipulation, and provenance is derived by comparing these two versions.  
This design, however, incurs significant memory overhead (since both versions must be stored in memory) and high computational costs when comparing them, especially for complex operations such as joins.  
By contrast, our approach is aware of the semantics of each data manipulation operation.  
For instance, in the case of a filter operation, we leverage the fact that the \texttt{DataFrame} index is preserved, allowing us to directly infer the lineage between input and output without storing or comparing full data records.  
For join operations, we retain the input \texttt{DataFrames} in memory and introduce unique record identifiers that can be used to efficiently infer provenance after the join, avoiding costly record comparisons, an optimization that our experiments show to be substantially more efficient.

The solution we have presented, in this paper, raises several questions:
\begin{itemize}[leftmargin=*, itemsep=0pt, topsep=0pt]
    \item Are tensors, as we have described them, an efficient method in terms of memory requirements for indexing the provenance of data preparation pipelines?
    \item What is the overhead in terms of processing time for capturing provenance information?
    \item How efficient is querying provenance using tensors?
    \item How does our solution compare with the state-of-the-art?
\end{itemize}

To answer the questions raised, we conducted empirical studies. 
For our experiments, we worked with three real-world data pipelines that involve various preprocessing steps. Table \ref{tab:peipelines} outlines the characteristics of the three pipelines in terms of the data processing steps they involve and the size of the datasets they manipulate. 
The German Credit pipeline is used to predict whether an individual is a suitable candidate for a loan. The Compas Score pipeline focuses on predicting an individual’s risk of recidivism. The third and final pipeline, viz the Census pipeline, aims to determine whether a person’s annual income exceeds \$50,000. For comparison with the state of the art, we considered the proposal by Chapman {\em et al.} as it is the only one sharing our objectives: (i) focusing on data preparation pipelines, and (ii) capturing their provenance in a queryable form.

\begin{table}[]
\footnotesize
\caption{Characteristics of the use case pipelines.}
\label{tab:peipelines}
\vspace{-0.25cm}
\begin{tabular}{l|c|c|c|c|c|}
\cline{2-6}
                                             & \textbf{\# Ops} & \textbf{\begin{tabular}[c]{@{}c@{}}\# Input \\ records\end{tabular}} & \textbf{\begin{tabular}[c]{@{}c@{}}\# Input \\ attr.\end{tabular}} & \textbf{\begin{tabular}[c]{@{}c@{}}\# Output \\ records\end{tabular}} & \textbf{\begin{tabular}[c]{@{}c@{}}\# Output \\ attr.\end{tabular}} \\ \hline
\multicolumn{1}{|l|}{\textbf{German}} & 4                     & 1000                                                               & 21                                                                     & 1000                                                                 & 60                                                                      \\ \hline
\multicolumn{1}{|l|}{\textbf{Compas}}  & 7                     & 7214                                                               & 53                                                                     & 6907                                                                 & 8                                                                       \\ \hline
\multicolumn{1}{|l|}{\textbf{Census}}        & 5                     & 32561                                                              & 15                                                                     & 32561                                                                & 104                                                                     \\ \hline
\end{tabular}
\vspace{-0.2cm}
\end{table}

\subsection{Memory Requirement for Indexing Provenance}
Table \ref{tab:memory-usecases} shows the memory requirements for storing the provenance of each use case using our proposed approach, TensProv, compared to the method proposed by Chapman {\em et al.} It demonstrates that our solution requires smaller memory compared with the proposal by  Chapman {\em et al.}. For example, in the census pipeline, which processes a relatively large dataset, 
the memory required by our TensProv is orders of magnitude less than that required by the Chapman et {\em al.}'s solution, corresponding to about 2.5 orders of magnitude less. This significant difference highlights the efficiency of our approach in terms of memory requirement.

\begin{table}[]
\vspace{-0.25cm}
\caption{Memory required for storing provenance using 
}
\label{tab:memory-usecases}
\centering
\vspace{-0.25cm}
\begin{tabular}{|l|c|c|c|}
\hline
 & \textbf{German} & \textbf{Compass} & \textbf{Census} \\ \hline
\textbf{TensorProv}   & 0.36 MB  & 3.52 MB   & 10.44 MB  \\ \hline
\textbf{Chapman et al.}    & 187.38 MB & 218.98 MB & 4012.38 MB \\ \hline
\end{tabular}
\vspace{-0.4cm}
\end{table}

\subsection{Provenance Capture Overhead}
\begin{figure}[]
\centering
\vspace{-0cm}
\includegraphics[width=0.5\textwidth]{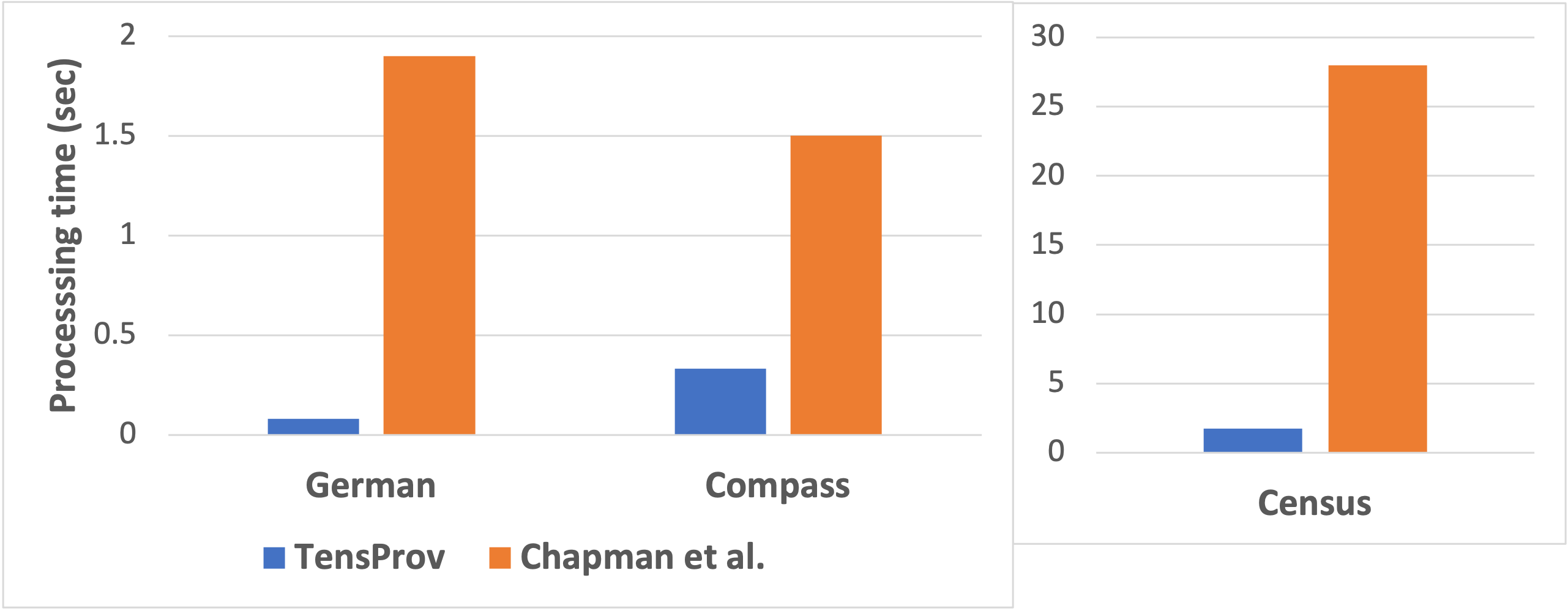}
\vspace{-0.7cm}
\caption{Overhead time due to provenance capture in the three uses cases.}
\label{fig:ovrehead-use-cases}
\vspace{-0cm}
\end{figure}

Figure~\ref{fig:ovrehead-use-cases} shows the overhead time for capturing provenance for the three  use cases using TensProv and the approach by Chapman~{\em et~al.}. 
TensProv incurs minimal processing overhead, even for the Census use case, which handles a larger dataset than the German and Compas use cases. Provenance capture stays below two seconds in all three use cases. When compared to the overhead of Chapman~{\em et~al.}’s solution, TensProv proves significantly more efficient. To quantify this improvement, we computed the speedup of TensProv relative to Chapman~{\em et~al.}’s approach (see Table~\ref{tab:speedup}). The results show that TensProv achieves substantial efficiency gains. These results stem from TensProv’s capture approach, which leverages data-processing semantics and dataframe indexes (when possible) to infer provenance without directly comparing dataframe contents before and after manipulation.

\begin{table}[]
\centering
\vspace{-0.2cm}
\caption{Speedup of TensProv compared with Chapman {\em et al.}}
\label{tab:speedup}
\vspace{-0.3cm}
\begin{tabular}{|l|l|l|l|}
\hline
\textbf{Use case} & \textbf{German} & \textbf{Compass} & \textbf{Census} \\ \hline
\textbf{Speedup}  & 23.75            & 4.55           & 16.18            \\ \hline
\end{tabular}
\end{table}

\subsection{Querying Provenance Information}

\begin{figure}[]
\centering
\vspace{-0.2cm}
\includegraphics[width=0.45\textwidth]{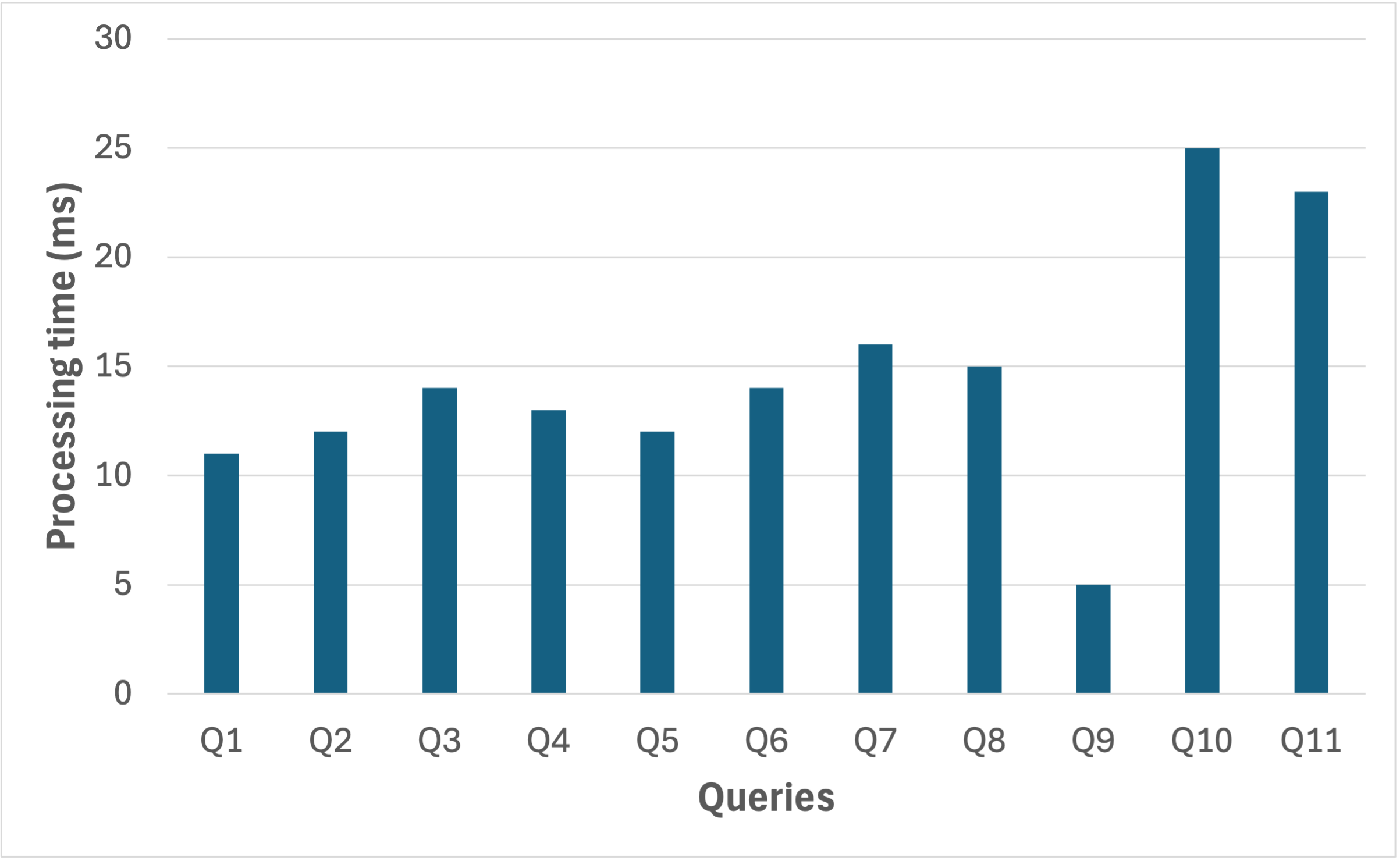}
\vspace{-0.5cm}
\caption{The processing time for each provenance query shown in Table \ref{tab:provenance_queries}.}
\label{fig:processingtime}
\vspace{-0.55cm}
\end{figure}

To evaluate performance, we executed all query types listed in Table 2 on the Census dataset (we modified the Census dataset to include join processing and cater for co-contribution and co-dependency queries, i.e. Q10 and Q11). Each query was run three times, and we report the average execution time. Figure~\ref{fig:processingtime} presents the results, showing that all queries are processed quickly, within the order of milliseconds.
Query 9 runs over the entire dataset and is particularly fast. This is because it does not require inspecting the tensor content. Only the associated metadata is needed to determine the type of operation performed. Queries 1 through 8 operate over data records or attribute values, and their execution times are relatively similar. This indicates that the bitset-based mechanisms we use to map from data records to attribute-value granularity perform efficiently. The last two queries (Q10 and Q11) take slightly longer than the others, as they involve both forward and backward traversals to identify co-contribution and co-influence. Yet, they are still evaluated within milliseconds.

\begin{figure}[h]
\centering
\vspace{-0.4cm}
\includegraphics[width=0.45\textwidth]{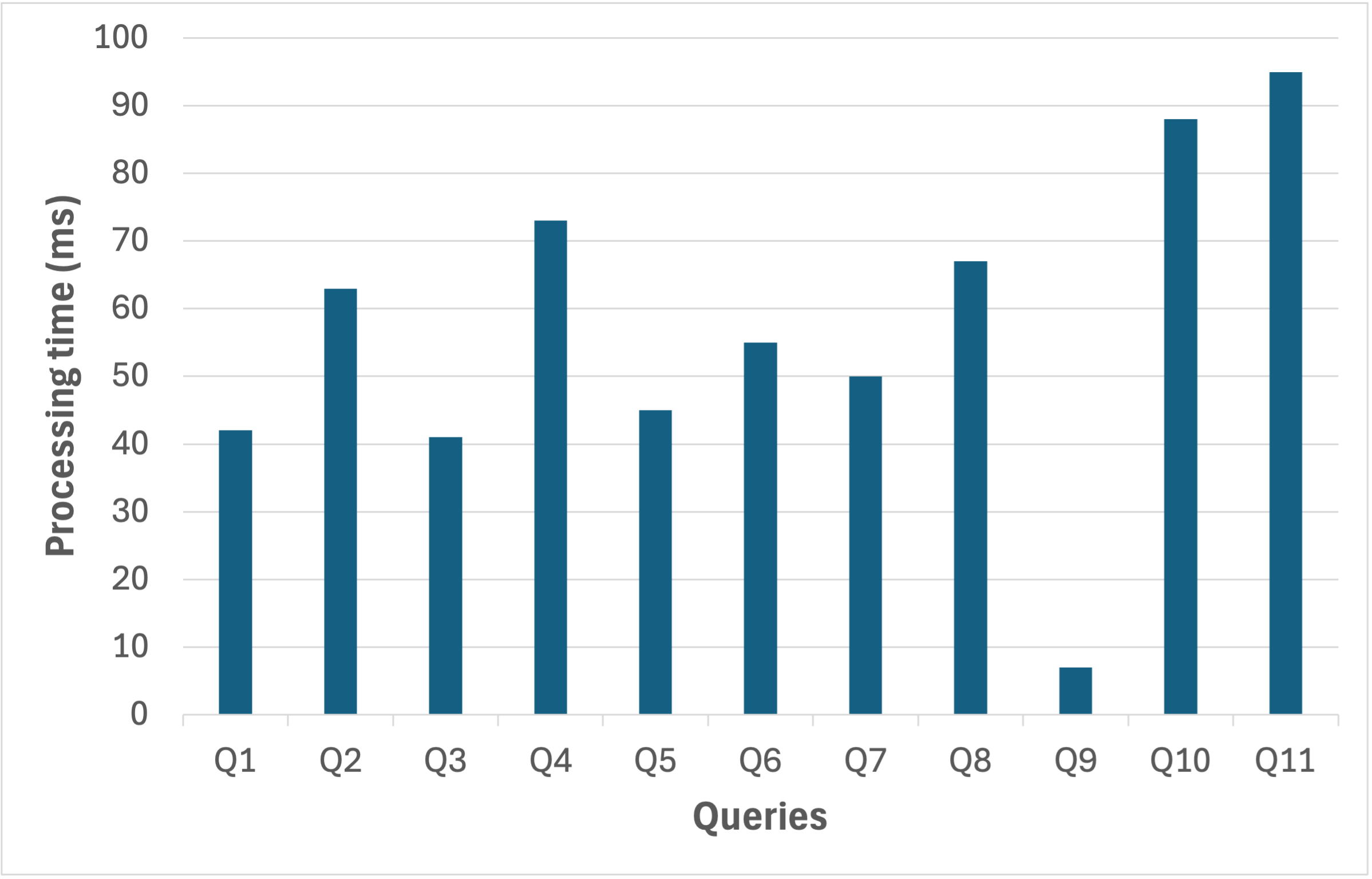}
\vspace{-0.5cm}
\caption{The processing time for queries when necessitating recomputation}
\label{fig:recomputationtime}
\vspace{-0.4cm}
\end{figure}

The provenance queries discussed above establish relationships between data records (or attributes) in the pipeline's input and output datasets. As these queries operate on materialized data, they require no recomputation. We further evaluated queries retrieving data from intermediate pipeline datasets (see Section~\ref{sec:localvscontextual}), which are non-materialized and thus necessitate recomputation.
Our experiments using the Census data (with queries detailed in Table~\ref{tab:provenance_queries}) measured performance when returning records from non-materialized intermediate dataframes. Figure~\ref{fig:recomputationtime} shows that query processing takes longer for non-materialized datasets compared to materialized ones (see Figure~\ref{fig:processingtime}). However, processing remains efficient, typically completing within milliseconds. This efficiency stems from recomputing only specific data records rather than entire datasets.
We observed  performance differences between forward and backward provenance queries. Forward queries (Q1, Q3, Q5, Q7) execute faster than their backward counterparts (Q2, Q4, Q6, Q8). This can be explained by the fact that backward queries require an additional back-tracing step to identify input records needed for recomputation. Query Q9 remains unaffected as it returns only transformation lists without data. Both co-contribution and co-influence queries show performance impacts from recomputation.


\subsection{In-Depth Examination of the Join }
We analyzed the behavior of individual data processing operators, focusing on the join operator, which is particularly demanding in terms of provenance capture. While operations such as {\em Filter} rely on dataframe indices to derive provenance, {\em Join} cannot exploit these indices and are therefore more computationally intensive. To assess the performance of our solution and compare it with that of Chapman~{\em et~al.}, we used the TPC-DI benchmark to generate synthetic datasets at scale factors of 3, 5, 9, 15, and 20.

We measured the memory required to encode the provenance using our approach (TensProv) and that of Chapman {\em et al.}. Table \ref{tab:join-provenance} summarizes, for each scale factor, the size of the datasets being joined (in terms of the number of records), the size of the provenance of the join results using TensProv, and the size of the provenance captured by Chapman {\em et al.}.
As shown in the table, the size of the provenance captured by our approach remains compact even as the size of the input datasets increases. Moreover, the provenance size produced by TensProv is orders of magnitude smaller compared to that of Chapman {\em et al.} For example, with a scale factor of 9, the provenance size using TensProv is approximately $29\text{MB}$, whereas it exceeds $3\text{GB}$ with the method proposed by Chapman {\em et al.}. Note also, that the approach by Chapman et al. does not scale, indeed, we were not able to process the join for scale factors of 15 and 20.

\begin{figure}[h]
\vspace{-0.2cm}
\centering
\includegraphics[width=0.5\textwidth]{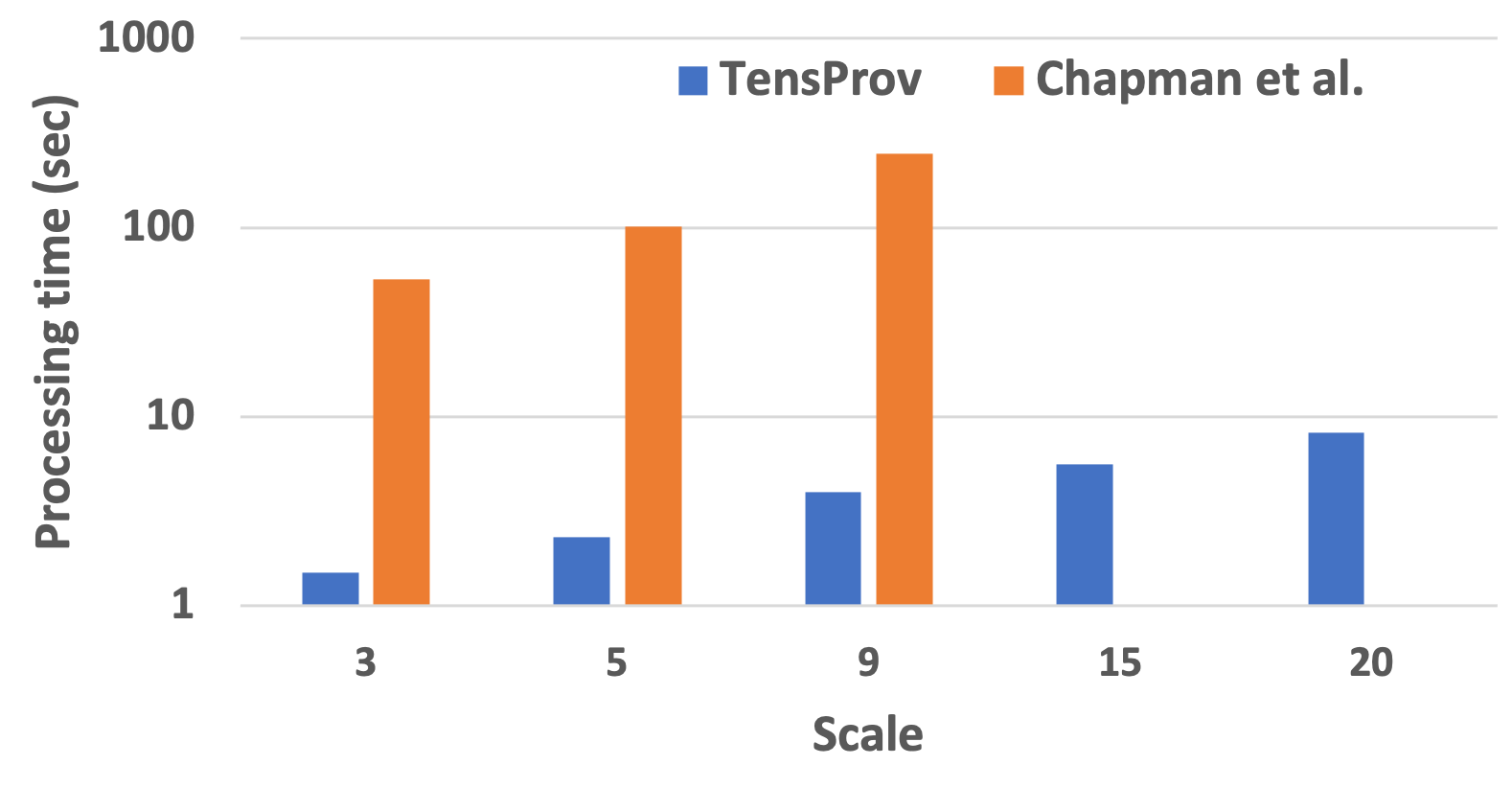}
\vspace{-0.8cm}
\caption{Processing time of the provenance of the join operator using TPC-DI.}
\label{fig:TDCIprocessingtime}
\vspace{-0.2cm}
\end{figure}

We also measured the time to capture provenance for the join operation (see Figure~\ref{fig:TDCIprocessingtime}). Evaluation time increases only slightly at larger scales (15 and 20) and remains much lower than for Chapman~{\em et~al.} This is because their inference-based employed by Chapman et al., even with hashing, is more costly, compared to TensProv, which captures provenance directly by instrumenting ID columns that are introduced for thi purpose in the input dataframes.

Query processing time, we measured the time to process why-provenance queries on the provenance captured for the join. The results show that evaluation time remains stable and insensitive to join size, with an average of approximately 0.04 seconds across all scale factors.

\begin{table}[]
\caption{Provenance size of the join operator using TPC-DI benchmark.}
\label{tab:join-provenance}
\vspace{-0.25cm}
\centering
\begin{tabular}{|c|c|c|c|}
\hline
\textbf{Scale factor} & \textbf{Datasets} & \textbf{TensProv} & \textbf{Chapman et al.} \\ \hline
3 & 362342\textbf{/}390978 & 3.02MB & 1.02GB \\ \hline
5 & 602956\textbf{/}650412 & 3.61MB & 1.78GB \\ \hline
9 & 1085239\textbf{/}1171107 & 6.50MB & 3.04GB \\ \hline
15 & 1807703\textbf{/}1951236 & 10.90MB & - \\ \hline
20 & 2411006\textbf{/}2601648 & 14.58MB & - \\ \hline
\end{tabular}
\vspace{-0.2cm}
\end{table}

In summary, our evaluation exercises demonstrated the effectiveness and efficiency of TensProv for provenance capture, both in processing and memory usage. It also showed that the captured provenance can be queried efficiently.

\section{Related Work}
\label{sec:relatedwork}

The capture and utilization of provenance have been extensively studied across various fields, including databases \cite{DBLP:journals/ftdb/CheneyCT09}, scientific workflows \cite{DBLP:conf/ipaw/PimentelDMBKMBL16}, and data science pipelines \cite{DBLP:conf/adbis/SchlegelS23}, which form the core focus of this paper. Focusing on the last domain, solutions like MLflow2Prov \cite{DBLP:conf/adbis/SchlegelS23} and Vamsa \cite{DBLP:conf/kdd/NamakiFPKAWZW20} address some challenges by capturing provenance data related to pipeline design. However, the provenance they generate primarily includes details such as the invocation of specific ML libraries through automated script analysis, rather than tracking data transformations. 

Several other platforms have a similar purpose, in that they aim  to capture broad metadata on ML pipeline (e.g., \cite{Schelter2018,DBLP:conf/adbis/SchlegelS23,DBLP:journals/debu/0001D18}). These platforms primarily focus on recording hyperparameter information, either extracted automatically or explicitly provided by ML pipeline developers. In terms of provenance, these systems typically offer coarse-grained tracking, identifying the datasets used and produced as well as the transformations applied to them. This is exemplified by ProvDB \cite{DBLP:journals/debu/0001D18}, which captures provenance implicitly by monitoring Unix-based commands. The provenance graph is stored in a graph database (e.g., Neo4J) and can be queried to determine which dataset version was used for training and with which hyperparameters. MLdp \cite{DBLP:conf/sigmod/AgrawalABBGGLMP19} integrates data management into the machine learning pipeline, aiming to handle the datasets used and manipulated during the training of ML models. 

Other tools are tailored to debugging ML pipelines. For instance, BugDoc \cite{DBLP:journals/vldb/LourencoFSWS23} identifies changes in preprocessing steps that result in model failures by analyzing script-level configurations and their sequences. Similarly, MLInspect \cite{DBLP:journals/vldb/GrafbergerGSS22} leverages provenance to detect data distribution shifts between the pipeline’s input and output, focusing on attributes like gender and ethnicity. This can be crucial for pinpointing biases introduced in the ML models trained using the output of the pipeline. In contrast to the above proposals, our work focuses on systematically capturing data provenance throughout pipeline processing and ensuring its seamless accessibility for querying and downstream exploitation.


Ursprung \cite{DBLP:journals/pvldb/RupprechtDAGB20,DBLP:conf/sigmod/RupprechtDALTB19} combines system-level provenance, such as call notifications, with application-specific sources, including log files and databases, to construct the provenance of ML pipelines. This integration enables users to explore various aspects of provenance, such as the transformations applied to a file, details about its provenance, and configuration parameters associated with the jobs (workloads) triggered by a pipeline execution. However, Ursprung does not seem to focus on capturing fine-grained provenance at the level of individual record values within a dataset. Instead, it relies on application-generated logs to construct provenance and emphasizes combining system-level and application-level provenance data to reconstruct the provenance of files (datasets as a whole), the transformations they underwent, and the resources, jobs, and configuration parameters used in the process.

The recent proposal by Chapman et al. \cite{DBLP:journals/tods/ChapmanLMT24} aims to capture the provenance of data pipelines and make it readily available for querying and exploitation. Our proposal differs in the following respects: (i) we provide an in-memory solution that allows pipeline developers to capture and query the provenance of the pipeline during its development. (ii) We do not attempt to capture the provenance of each cell (attribute-value) individually. As our experiments show, such an approach is resource-greedy in terms of both memory usage and provenance capture. Instead, we adopt an approach that captures provenance at the level of data records and augments this provenance with metadata for the inference of provenance information at the cell level.

It is worth mentioning Yan \emph{et al.}'s proposal for tracking tensor manipulation provenance \cite{10.5555/3026947.3026948}, which focuses on linear algebra operations like addition, multiplication, and inversion. In contrast, our approach uses tensors to encode fine-grained provenance for dataset manipulations, emphasizing data transformations rather than linear algebra computations.

\section {Conclusions}
\label{sec:conclusions}

In this work, we presented {\em TensProv}, an in-memory solution for capturing and querying the provenance of data preparation pipelines. Unlike disk-based solutions, our approach enables dynamic, real-time exploration of provenance during the pipeline development phase, providing developers with  feedback to identify and resolve issues. By leveraging sparse binary tensors, we demonstrated an efficient mechanism for recording record-level provenance and augmenting it with schematic metadata to infer fine-grained attribute-level dependencies. 
The evaluation exercises also demonstrate the efficiency of our solution in processing a broad range of provenance queries.



\bibliographystyle{plain}
\bibliography{references}

\end{document}